\documentclass[aps,prb,reprint,amsmath,amssymb,superscriptaddress,showpacs]{revtex4-1}
\usepackage{graphicx}
\usepackage{color}
\usepackage{hyperref}
\usepackage{multirow}

\newcommand{\tc}{\ensuremath{T_{\textrm C}}}
\newcommand{\mB}{\ensuremath{\mu_{\textrm B}}}

\begin{document}

\title{X-ray absorption spectroscopy study of the electronic and magnetic proximity effects in YBa$_2$Cu$_3$O$_{7}$/La$_{2/3}$Ca$_{1/3}$MnO$_3$ and La$_{2-x}$Sr$_x$CuO$_4$/La$_{2/3}$Ca$_{1/3}$MnO$_3$ multilayers}

\author{M.\,A.~\surname{Uribe-Laverde}}
\email{miguelangel.uribelaverde@unifr.ch}
\author{S.~Das}
\thanks{M.A.U.L. and S.D. contributed equally to this work.}
\author{K.~Sen}
\author{I.~Marozau}
\author{E.~Perret}
\affiliation{University of Fribourg, Department of Physics and Fribourg Centre for Nanomaterials, Chemin du Mus\'ee 3, CH-1700 Fribourg, Switzerland}

\author{A.~Alberca}
\affiliation{University of Fribourg, Department of Physics and Fribourg Centre for Nanomaterials, Chemin du Mus\'ee 3, CH-1700 Fribourg, Switzerland}
\affiliation{Swiss Light Source, Paul Scherrer Institut, CH-5232 Villigen PSI, Switzerland}

\author{J.~Heidler}
\author{C.~Piamonteze}
\affiliation{Swiss Light Source, Paul Scherrer Institut, CH-5232 Villigen PSI, Switzerland}

\author{M.~Merz}
\author{P.~Nagel}
\author{S.~Schuppler}
\affiliation{Karlsruhe Institute of Technology, Institut f\"ur Festk\"orperphysik, D-76021 Karlsruhe, Germany}

\author{D.~Munzar}
\affiliation{Department of Condensed Matter Physics, Faculty of Science, Masaryk University, Kotlarska 2, 61137 Brno, Czech Republic}

\author{C.~Bernhard}
\affiliation{University of Fribourg, Department of Physics and Fribourg Centre for Nanomaterials, Chemin du Mus\'ee 3, CH-1700 Fribourg, Switzerland}


\date{\today}

\begin{abstract}
With x-ray absorption spectroscopy we investigated the orbital reconstruction and the induced ferromagnetic moment of the interfacial Cu atoms in YBa$_2$Cu$_3$O$_{7}$/La$_{2/3}$Ca$_{1/3}$MnO$_3$ (YBCO/LCMO) and La$_{2-x}$Sr$_{x}$CuO$_4$/La$_{2/3}$Ca$_{1/3}$MnO$_3$ (LSCO/LCMO) multilayers. We demonstrate that these electronic and magnetic proximity effects are coupled and are common to these cuprate/manganite multilayers. Moreover, we show that they are closely linked to a specific interface termination with a direct Cu-O-Mn bond. We furthermore show that the intrinsic hole doping of the cuprate layers and the local strain due to the lattice mismatch between the cuprate and manganite layers are not of primary importance. These findings underline the central role of the covalent bonding at the cuprate/manganite interface in defining the spin-electronic properties.
\end{abstract}


\maketitle
\section{Introduction}
The new phenomena that arise when different interactions and orders meet at the interfaces of multilayers (MLs) from complex oxides with strongly correlated electrons are of great current interest~\cite{Hwang_2012}. The versatile properties of the complex oxides provide many opportunities, but they make it also very challenging to identify the underlying principles and interactions.

A prominent example are MLs from cuprate high \tc\ superconductors and ferromagnetic (FM) manganites. They have appealing properties, like the high superconducting (SC) critical temperatures, \tc, of the cuprates and the nearly 100\% spin polarized charge carriers of the manganites. Studies on YBa$_2$Cu$_3$O$_7$/La$_{2/3}$Ca$_{1/3}$MnO$_3$ (YBCO/LCMO) MLs have already provided evidence for a giant magneto resistance effect~\cite{Pena_2005}, a giant superconductivity induced change of the ferromagnetic order~\cite{Hoppler_2009} and a proximity-induced spin-triplet SC state~\cite{Visani_2012,Golod_2013}. The latter is a well-established phenomenon in conventional SC/FM MLs where it requires a specific magnetic structure with a non-collinear order near the interface~\cite{Bergeret_2005,Eschrig_2011,Keizer_2006,*Robinson_2010,*Khaire_2010}.

The corresponding magnetic properties of the cuprate/manganite MLs remain to be understood. The YBCO/LCMO MLs have been shown to exhibit an unusual magnetic proximity effect (MPE) that involves a suppression of the FM order of the Mn moments on the LCMO side of the interface~\cite{Stahn_2005,Satapathy_2012,Uribe_2013} and, yet, an induced FM moment of the Cu atoms on the YBCO side. The latter has been established with Cu L-edge x-ray magnetic circular dichroism (XMCD) studies~\cite{Chakhalian_2006,Satapathy_2012,Giblin_2012}. Notably, these Cu moments are antiparallel with respect to the Mn moments. A subsequent x-ray linear dichroism (XLD) study suggested that an orbital reconstruction of the interfacial Cu atoms and a charge transfer across the interface are at the heart of this MPE~\cite{Chakhalian_2007}. The orbital reconstruction, the charge transfer and the antiferromagnetic (AF) coupling between Cu and Mn moments have been explained in terms of a covalent bonding of Cu and Mn via the apical oxygen, O$^{\textrm{ap}}$~\cite{Chakhalian_2007}. This covalent bonding model relies on the specific interface termination of these YBCO/LCMO MLs with a CuO$_2$-BaO-MnO$_2$ layer stacking which has been observed with high resolution transmission electron microscopy (HRTEM)~\cite{Varela_2003,Malik_2012}.

The relevance of the covalent bonding model and, in particular, how these electronic and magnetic proximity effects depend on the interface termination, on the local strain and subsequent defects and on the hole doping and thus the strength of the magnetic correlations of the cuprate layers need to be explored. This is especially important since subsequent experimental and theoretical studies on YBCO/LCMO MLs yielded controversial results which question the direct relationship between the orbital reconstruction and the induced Cu moments~\cite{Werner_2010,Yang_2009}.

In this paper we present a combined XLD and XMCD study on YBa$_2$Cu$_3$O$_{7}$/La$_{2/3}$Ca$_{1/3}$MnO$_3$ (YBCO/LCMO) and La$_{2-x}$Sr$_{x}$CuO$_4$/La$_{2/3}$Ca$_{1/3}$MnO$_3$ (LSCO/LCMO) MLs which establishes a direct relationship between the orbital reconstruction and the induced FM moment of the interfacial Cu atoms and underlines the important role of the interface termination. The unique properties of the LSCO/LCMO system also allow us to vary the hole doping of the CuO$_2$ planes and to explore the role of the local strain due to the lattice mismatch between the cuprate and manganite layers.

\section{Experimental}\label{exp}
The pulsed laser deposition (PLD) growth of the LSCO/LCMO [10\,nm/10\,nm]$_{\times 9}$ and [8.5\,nm/8.5\,nm]$_{\times 3}$ (9\_BL and 3\_BL) MLs on SrLaAlO$_4$ (SLAO) substrates has been reported in Ref.~\onlinecite{Das_2014}. The one of a corresponding YBCO/LCMO [10\,nm/10\,nm]$_{\times 10}$ ML on LSAT in Ref.~\onlinecite{Malik_2012}.

The XMCD and XLD studies were performed at the XTreme beamline at the Swiss Synchrotron Light Source (SLS) of Paul Scherrer Institute (PSI) in Villigen, Switzerland~\cite{Piamonteze_2012}. Additional measurements (not shown) were done at the WERA beamline of the ANKA synchrotron at K.I.T., Germany. The x-ray absorption spectroscopy (XAS) curves were measured at the Mn and Cu-L$_{3,2}$ edges using the total electron yield (TEY) and total fluorescence yield (FY) modes. The data have been normalized and background-corrected following the procedure detailed in Appendix~\ref{Back_corr}.

For XLD the absorption for $\sigma$ and $\pi$ linear polarization was measured with an angle of incidence of 15$^{\circ}$ for YBCO/LCMO and LSCO/LCMO 9\_BL and 20$^{\circ}$ for LSCO/LCMO 3\_BL. The absorption for polarization parallel to the planes, $\mu_{ab}$, is directly obtained from the $\sigma$ polarization measurement. To obtain the absorption for the out-of-plane polarization, $\mu_c$, the measurement geometry needs to be considered since in $\pi$ polarization the electric field has finite in-plane and out-of-plane components. As a function of the incidence angle with respect to the sample surface, $\theta$, the absorption for $\pi$ polarization yields~\cite{Stohr_1995}: $\mu_{\pi}=\mu_{ab}\sin^2\theta+\mu_{c}\cos^2\theta$. The value of  $\mu_c$ can thus be calculated as:
\begin{equation}
\mu_{c}=\frac{\mu_{\pi}-\mu_{\sigma}\sin^2\theta}{\cos^2\theta}.
\end{equation}
For XMCD the absorption for circular polarization with positive and negative helicity, $\mu_+$ and $\mu_-$, was measured with the same incident angles as for XLD. The magnetization was always first saturated at 3\,T before the XMCD was measured at 0.5\,T applied along the direction of the incident x-rays. The dichroism has been obtained by switching both the direction of the applied field and the x-ray polarization.
\begin{figure}[tb]
\centering
\includegraphics[width=0.45\textwidth]{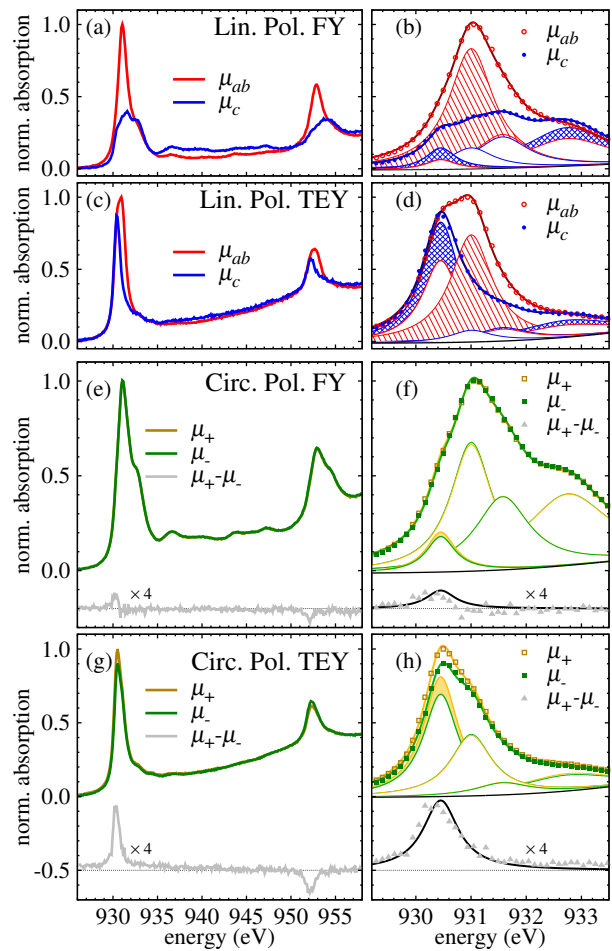}
\caption{(Color online) XAS curves at the Cu-L$_{3,2}$ edge of YBCO/LCMO taken at 2\,K. (a) XAS data for linear polarization in FY mode. (b) Close-up of the L$_3$ edge. Data (points) are shown together with the fit (thick lines) using four Lorentzian functions (thin lines) and a background with a linear and a sigmoid function (thin black line). The XLD of each peak is indicated by the patterns. (c) and (d) Corresponding XAS data for linear polarization in TEY mode. (e) XAS and XMCD data for circular polarization in FY mode. (f) Close-up of the L$_3$ edge shown with the same symbols as in (b). Also shown is the fit of the XMCD signal (black line). (g) and (h) Corresponding XAS and XMCD data for circular polarization in TEY mode.\label{410}}
\end{figure}
\begin{figure}[tb]
\centering
\includegraphics[width=0.45\textwidth]{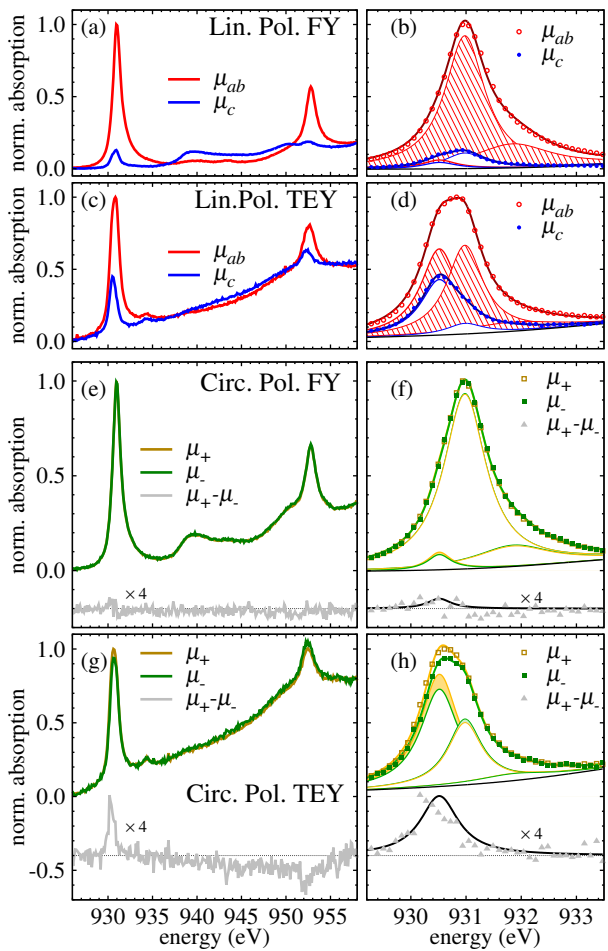}
\caption{(Color online) XAS curves at the Cu-L$_{3,2}$ edge of LSCO/LCMO 9\_BL at 2\,K shown with the same symbols as in Fig.~\ref{410}.\label{1201}}
\end{figure}

\section{Results and discussion}
Figures~\ref{410} and ~\ref{1201} compare the Cu L-edge XAS curves, measured in the FY and TEY modes with linear and circular polarization, of the YBCO/LCMO and LSCO/LCMO MLs, respectively. They show that both MLs exhibit clear signatures of an orbital reconstruction and an induced FM moment of the interfacial Cu atoms. For the following discussion it is important to recall that the TEY mode has a very small probe depth of only a few nanometers which is limited by the escape depth of the photoelectrons. The TEY signal is thus governed by the response of the Cu atoms at the topmost cuprate/LCMO interface. The FY mode has a much larger probe depth of about 100\,nm and thus yields an averaged response of the Cu atoms in several cuprate layers.

The orbital reconstruction of the interfacial Cu atoms is evident from the different polarization dependence of the XAS curves in FY and TEY mode in Figs.~\ref{410}(a)-\ref{410}(d) for YBCO/LCMO and~\ref{1201}(a)-\ref{1201}(d) for LSCO/LCMO. The FY curves yield a strong dependence of the absorption on the polarization direction that is characteristic of the bulk cuprates. Due to the predominant d$_{x^2-y^2}$ character of the Cu holes in the CuO$_2$ layers, the absorption peaks at the Cu-L$_{3,2}$ edges are much more intense for the in-plane polarization~\cite{Fink_1994,Nuecker_1995}. This so-called XLD effect is much weaker in the TEY curves for which the absorption peaks for the out-of-plane polarization also exhibit a sizable intensity. This observation suggests that for the Cu atoms at the topmost interface the holes are more equally distributed between the d$_{x^2-y^2}$ and d$_{3z^2-r^2}$ levels. This orbital reconstruction was previously reported for YBCO/LCMO~\cite{Chakhalian_2007}, our new data show that it occurs equally well for LSCO/LCMO and thus is common to the cuprate/manganite MLs.
\begin{table}
  \centering
  \begin{tabular}{|c|c|c|c|c|}\hline
  \multirow{2}{*}{Peak}   &  \multirow{2}{*}{Mode}   & \multirow{2}{*}{YBCO/LCMO} & LSCO/LCMO & LSCO/LCMO \\
      &   &          & 9\_BL & 3\_BL \\\hline
\multicolumn{1}{ |c| }{\multirow{2}{*}{1}} & FY & 4.7(7) & 1.9(7) &  -- \\
                       \multicolumn{1}{ |c| }{} & TEY & 41(3)  & 48(5) & 14(1) \\\hline
\multicolumn{1}{ |c| }{\multirow{2}{*}{2}} & FY & 45(2)  & 73(3) & -- \\
                       \multicolumn{1}{ |c| }{} & TEY & 37(4)  & 39(5) & 62(4) \\\hline
\multicolumn{1}{ |c| }{\multirow{2}{*}{3}} & FY & 20(2)  & 25(3) & -- \\
                       \multicolumn{1}{ |c| }{} & TEY & 8(2)   & 13(8) & 24(3) \\\hline
\multicolumn{1}{ |c| }{\multirow{2}{*}{4}} & FY & 31(3)  & -- & -- \\
                       \multicolumn{1}{ |c| }{} & TEY & 15(4)  & -- & -- \\\hline
  \end{tabular}
  \caption{Normalized weight in the polarization-averaged spectra, $(2\mu_{ab}+\mu_c)/3$, of the peaks contributing to the Cu-L$_3$ absorption in FY and TEY modes. The peaks are labeled from lowest to highest energy. All the values are given in percentages.\label{results}}
\end{table}

A quantitative analysis of the electric and magnetic properties of the Cu atoms at the interface has been carried out by means of a multi-component peak fit of the absorption spectra in the region around the Cu-L$_3$ edge. The resonant signal has been fitted with up to four Lorentzian functions. The background and the edge jumps in the vicinity of the absorption edge have been accounted for with linear and sigmoid functions. To minimize the number of free parameters, we have used for each of the Lorentzian functions the same peak energy to fit simultaneously all of the measured FY and TEY curves of a given sample. The widths of the Lorentzian peaks has also been constrained to a common value that was however allowed to vary between the FY and TEY curves.

The multiplet fitting of the absorption peaks at the L$_3$-edge shown in Figs.~\ref{410}(b,d,f,h) and~\ref{1201}(b,d,f,h) reveals a clear substructure that is similar for the YBCO/LCMO and LSCO/LCMO MLs. In the former, it consists of four subpeaks at about 930.5, 931, 931.6 and 932.8\,eV. The normalized weight, $w$, of each peak in the polarization-averaged absorption curve, $(2\mu_{ab} + \mu_{c})/3$, has been calculated from the fitted intensities and is listed in Table~\ref{results}. As detailed in Appendix~\ref{weight}, the weight of each contribution is proportional to the fraction of probed Cu atoms for which the corresponding transition occurs. In FY mode, which probes the Cu atoms of several YBCO layers, $w$ corresponds to the percentage of Cu atoms in the sample contributing to the transition. In TEY mode such a quantitative interpretation is not possible due to the unknown electron escape depth. The comparison of the relative weights of the peaks in FY and TEY mode still allows us to distinguish between the transitions from Cu atoms that are located right at the interface or away from it. For example, the peak at 930.5\,eV has a much larger weight in the TEY curves than in the FY curves which indicates that it originates from Cu atoms at the YBCO/LCMO interfaces. The intensity of this peak is larger for the out-of-plane polarization, the Cu atoms at the interface where this transition takes place are therefore orbitally reconstructed and their holes have a predominant d$_{3z^2-r^2}$ character.

The other peaks can be assigned to Cu atoms away from the interface since they are stronger in FY mode and similar peaks occur in bulk YBCO~\cite{Nuecker_1995}. The peak at 931\,eV corresponds to the 2p$^6$3d$^9$$\rightarrow$2p$^5$d$^{10}$ transition. As already mentioned, the absorption intensity of this peak for the out-of-plane polarization is small due to the predominant in-plane character of holes in the CuO$_2$ planes. The other two peaks have been assigned to 2p$^6$3d$^9$\underline{L}$\rightarrow$2p$^5$d$^{10}$\underline{L} transitions, where \underline{L} denotes a hole on the oxygen ligand. Their higher energy arises from the Coulomb repulsion due to the overlap between the core hole and the oxygen ligand hole. The 3d$^9$\underline{L} state corresponds to the so-called \textquotedblleft Zhang-Rice singlet\textquotedblright in the CuO$_2$ planes and CuO chains. In the latter, they have a higher concentration and a strong out-of-plane character~\cite{Nuecker_1995}. This explains the similar intensity of the in-plane and out-of-plane absorption of the peaks at 931.6 and 932.8\,eV.

Notably, the XAS curves of the LSCO/LCMO ML in Figs.~\ref{1201}(b,d,f,h) show a corresponding low-energy peak at 930.5 eV that is also much stronger in TEY than in FY mode and thus can be assigned to the Cu atoms at the LSCO/LCMO interfaces. The orbital reconstruction of these atoms is confirmed by the reduced difference between the in-plane and out-of-plane polarization as compared with the bulk behavior. The fact that the absorption intensity for the in-plane polarization is still larger than the out-of-plane suggest that the orbital reconstruction effect is somewhat weaker than in the YBCO/LCMO ML. The peaks at 931\,eV and 931.8\,eV resemble the ones in bulk LSCO for which they are associated with the 2p$^6$3d$^9$$\rightarrow$2p$^5$d$^{10}$ and the 2p$^6$3d$^9$\underline{L}$\rightarrow$2p$^5$d$^{10}$\underline{L} transitions, respectively~\cite{Fink_1994}. Since there are no CuO chains in LSCO, the holes and Zhang-Rice singlets reside only in the planes and there is no fourth peak around 932.5 eV.

\begin{figure}[tb]
\centering
\includegraphics[width=0.45\textwidth]{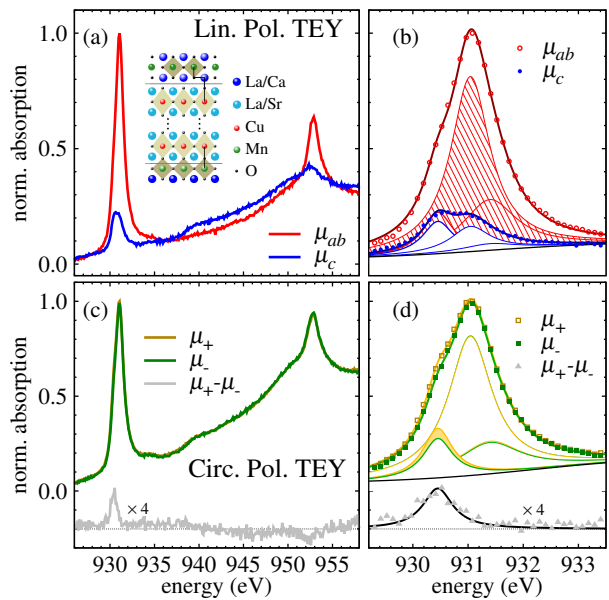}
\caption{(Color online) XAS curves at the Cu-L$_{3,2}$ edge of LSCO/LCMO 3\_BL at 2\,K in the TEY mode shown with the same symbols as in Fig.~\ref{410}. Inset of panel~(a): sketch of the different LSCO/LCMO interface terminations. The bottom interface exemplifies the direct Cu-O$^{\textrm{ap}}$-Mn bond, the top one displays the zig-zag-type Cu-O$^{\textrm{ap}}$-La-O$^{\textrm{ap}}$-Mn bond.\label{1337}}
\end{figure}

A significant difference between the YBCO/LCMO and LSCO/LCMO MLs concerns the weight of the 930.5\,eV peak in the FY curves which amounts to about 4.7\% and 2\%, respectively (see Table~\ref{results}). While the absolute value of $w$ should be considered with care, since it depends on the normalization and the background subtraction, the relative difference is still meaningful. It indicates that the fraction of orbitally reconstructed Cu atoms in LSCO/LCMO is significantly smaller than in YBCO/LCMO. This can indeed be understood in terms of the different interface termination. For YBCO/LCMO the top and bottom interfaces are reported to have the same termination with a straight Cu-O$^{\textrm{ap}}$-Mn bond (we call it type-I termination)~\cite{Varela_2003,Malik_2012} that gives rise to the covalent bonding and the orbital reconstruction. For LSCO/LCMO, as sketched in the inset of Fig.~\ref{1337}(a), there exists a second kind of termination with a zig-zag-type Cu-O$^{\textrm{ap}}$-La-O$^{\textrm{ap}}$-Mn bond~\cite{Biskup_2014} which should occur with equal probability. This so-called type-II termination yields a much weaker covalent bonding between the Cu and Mn ions which is not expected to support the orbital reconstruction.

The important role of the interface termination for the orbital reconstruction phenomenon is evident from the comparison of the TEY curves of the 9\_BL and 3\_BL LSCO/LCMO samples in Figs.~\ref{1201}(c,d) and~\ref{1337}(a,b), respectively. TEM studies on these LSCO/LCMO MLs have shown that the termination at the SLAO/LSCO interface is predominantly of type-I with Al-O$^{\textrm{ap}}$-Cu bonds~\cite{Das_2014,Biskup_2014}. This result is rather surprising since the termination of the SLAO substrate has not been actively controlled. Nevertheless, we note that a similar effect in terms of an atomically precise control of an interface in the presence of non-stoichiometric deposition conditions has been reported in Ref.~\onlinecite{Nie_2014}. As a consequence, the termination at the subsequent first LSCO/LCMO interface is preferably of type-II. For the following LSCO/LCMO bilayers the regularity of this asymmetric termination pattern is eventually lost, likely due to strain-induced defects that originate from the sizable lattice mismatch between LSCO and LCMO. The fraction of type-I termination at the topmost LSCO/LCMO interface that is probed in TEY mode is therefore expected to be significantly larger for the 9\_BL than for the 3\_BL sample. In good agreement, we find that the weight of the 930.5\,eV peak in the TEY curves is about 3 times larger for the 9\_BL than for the 3\_BL sample (see Table~\ref{results}).

Next, we discuss the relationship between the orbital reconstruction and the induced FM moment of interfacial Cu. As shown in Figs.~\ref{410}(e,g), \ref{1201}(e,g) and~\ref{1337}(c), the YBCO/LCMO and LSCO/LCMO MLs exhibit a finite XMCD signal, defined as $\mu_+-\mu_-$.  The opposite sign of the XMCD at the L$_3$ and L$_2$ edges evidences that it has a magnetic origin. The larger XMCD intensity in the TEY mode as compared with the FY mode confirms that the FM Cu moments are located at the interfaces. The direct link between the orbital reconstruction and the induced FM in Cu becomes evident from the close-up of the absorption curves at the L$_3$ edge. The fits of the XAS curves for YBCO/LCMO in Figs.~\ref{410}(f,h) and for LSCO/LCMO in Figs.~\ref{1201}(f,h) and~\ref{1337}(d) show indeed that only the low-energy peak at 930.5\,eV exhibits a sizable XMCD. This important finding highlights that the FM moment originates from the very same Cu atoms which undergo the orbital reconstruction. A direct link between the electronic and magnetic proximity effects is thus confirmed. The sign of the XMCD signal is the same for LSCO/LCMO and YBCO/LCMO and indicates that the Cu moments are always antiparallel to the Mn moments.
\begin{table}
\centering
\begin{tabular}{c|c|c|c|}\cline{2-4}
         & \multirow{2}{*}{YBCO/LCMO} & LSCO/LCMO & LSCO/LCMO \\
         &          & 9\_BL & 3\_BL \\\hline
\multicolumn{1}{ |c| }{$m_{s}^\textrm{Cu,int}$ ($\mu_{\textrm{B}}$)} & -0.18(4) & -0.16(4) & -0.15(4)\\\hline
\multicolumn{1}{ |c| }{$m_{s}^\textrm{Mn}$ ($\mu_{\textrm{B}}$)} & 1.7(3) & 2.0(4) & 2.6(5)\\\hline
\end{tabular}
\caption{Calculated Cu and Mn spin magnetic moments of the YBCO/LCMO and LSCO/LCMO MLs. The values are calculated applying the sum rules to the TEY XMCD data at the Cu and Mn-L$_{3,2}$ edges. The non-resonant background in the XAS curves has been subtracted using the approach described in Ref.~\onlinecite{Uribe_2014a}. For Cu, an additional normalization to the weight of the 930.5\,eV peak in the XAS curves has been applied to obtain the magnetization of the FM ordered atoms at the interface.\label{moments}}
\end{table}

Since the FM order of the Cu atoms is confined to the interface the fraction of probed FM Cu atoms in a XMCD experiment is generally unknown. Therefore, an accurate determination of their magnetization is commonly elusive~\cite{Chakhalian_2006,Werner_2010}. This difficulty is solved by the multiplet fitting analysis introduced in this paper. Since only the orbitally reconstructed atoms are magnetic, their fraction corresponds to the weight of the  930.5\,eV peak in the polarization averaged absorption shown in Table~\ref{results}. After performing the sum rule analysis described in Ref.~\onlinecite{Uribe_2014a} and a subsequent normalization to the fraction of orbitally reconstructed atoms we obtain the Cu magnetization values shown in Table~\ref{moments}. Notably, although the number of magnetic Cu atoms differs between these YBCO/LCMO and LSCO/LCMO MLs the magnetic moment does not exhibit a significant variation. Table~\ref{moments} also shows the calculated moment obtained from the XMCD measurements in TEY mode performed at the Mn-L$_{3,2}$ edge (spectra are not shown). They confirm  that the topmost LCMO layers of the YBCO/LCMO and LSCO/LCMO MLs exhibit a sizable FM order.

We also explored the role of the hole doping of the CuO$_2$ planes and thus of the strength of the antiferromagnetic (AF) correlations that are intrinsic to the cuprate layers. In YBCO it is controlled by the oxygen content of the CuO chains which act as charge reservoirs for the CuO$_2$ planes and at full oxygenation yield an optimally doped state ($p\approx 0.15$) with maximal~\tc. However, according to HRTEM studies, the CuO$_2$ planes next to the interface are lacking the neighboring CuO chain layer and thus may be strongly underdoped with static Cu moments and strong, intrinsic AF correlations. According to Refs.~\onlinecite{Varela_2003,Chakhalian_2007} the hole doping of the interfacial CuO$_2$ planes may be further reduced by a transfer of electrons from the manganite to the cuprate layers. In contrast, the LSCO/LCMO system allows one to determine and vary the intrinsic doping of the interfacial CuO$_2$ planes via the Sr substitution. Accordingly, we investigated additional LSCO/LCMO MLs with $x=0$ and $0.3$ for which the intrinsic doping yields an undoped state with long-range AF order or a strongly overdoped and essentially non magnetic state, respectively. Notably, as presented in Table~\ref{mCu_doping} our XMCD data (spectra are not shown) reveal that the magnitude of the FM moment per orbitally reconstructed Cu atom is almost independent of the Sr content and thus hardly depends on the intrinsic doping and the strength of the intrinsic AF correlations of the cuprate layers. This confirms that this MPE is governed by the exchange coupling due to the local covalent bonding between the Cu and Mn atoms and the possibly related transfer of electrons from the manganite to the cuprate layers.
\begin{table}
\centering
\begin{tabular}{|c|ccc|}\hline
 & x=0 & x=0.15 & x=0.30 \\\hline\hline
3\_BL & -0.16(3) & -0.15(4) & -0.16(4)\\
9\_BL & -0.20(4) & -0.14(4) & -0.22(5)\\\hline
\end{tabular}
\caption{Calculated spin magnetic moment, in $\mB$, per orbitally reconstructed Cu atom for the LSCO/LCMO multilayers with different Sr doping levels and bilayer repetitions. The calculations have been performed using the TEY XMCD data (not shown) following the procedure described in Ref.~\onlinecite{Uribe_2014a}.\label{mCu_doping}}
\end{table}

Finally, the comparison of LSCO/LCMO and YBCO/LCMO allowed us to explore the role of the local strain at the interface and subsequent defects. This is due to the different mismatch of the in-plane lattice parameters which amounts to about 2.5\% between LSCO and LCMO~\cite{Das_2014} and 0.5\% between YBCO and LCMO~\cite{Malik_2012}. The similar behavior of the LSCO/LCMO and YBCO/LCMO MLs as discussed above suggests that the orbital reconstruction and the induced FM moment of the interfacial Cu atoms are not strongly affected by these strain effects. Nevertheless, the magnitude of the orbital reconstruction is somewhat larger in YBCO/LCMO than in LSCO/LCMO. These observations should motivate calculations based on the covalent bonding model of Ref.~\onlinecite{Chakhalian_2007} to detail the relationship of the orbital reconstruction with the splitting and the relative shift of the $e_g$ levels of the interfacial Cu and Mn atoms and with the displacement of the apical oxygen along the Cu-O$^{\textrm{ap}}$-Mn bond~\cite{Yang_2009}. A related effect that deserves further theoretical attention involves the broader peaks above 935\,eV which according to Ref.~\onlinecite{Fink_1994} originate from transitions into hybridized Cu-4s and Cu-3d$_{3z^2-r^2}$ levels. Figures~\ref{410}(a,c) and~\ref{1201}(a,c) show that these are considerably weaker in the TEY than in FY curves which suggests a modification of the local structure around the interfacial Cu atoms, for example due to a change of the Jahn-Teller distortion or the buckling of the CuO$_2$ planes. Such distortions may provide a complementary or alternative explanation for the red shift of the 930.5\,eV peak that was previously interpreted in terms of an electron transfer from the manganite to the cuprate layers.
\section{Conclusions}
In summary, with x-ray absorption spectroscopy on LSCO/LCMO and YBCO/LCMO multilayers we have shown that the orbital reconstruction and the induced ferromagnetic moment of the interfacial Cu atoms are common to these cuprate/manganite multilayers. We have also established a direct link between these electronic and magnetic interface phenomena and we have shown that they require a specific kind of interface termination. The intrinsic hole doping of the cuprates and the local strain due to the lattice mismatch between the different layers is found to be less important. Our findings support the covalent bonding model~\cite{Chakhalian_2007} and related models~\cite{Salafranca_2010}. They also highlight the important role of the interface termination in defining the spin-electronic properties of these oxide multilayers.
\begin{acknowledgments}
The work at UniFr has been supported by the Schweizer Nationalfonds (SNF) grants 200020-140225 and 200020-153660. The XAS was measured at the EPFL/PSI X-Treme beamline of the Swiss Light Source at Paul Scherrer Institut, Villigen, Switzerland and the WERA beamline at the ANKA synchrotron, Karslruhe Institute of Technology, Germany. D.M. was supported by the CEITEC project (CZ.1.05/1.1.00/02.0068). We thank Mar\'{\i}a Varela and Neven Biskup for helpful discussions of their TEM data on the LSCO/LCMO MLs.
\end{acknowledgments}

\appendix

\section{Normalization and background subtraction of the TEY and FY absorption curves}\label{Back_corr}
To single out the resonant part of the absorption due to the Cu-L$_{3,2}$ edges, it was necessary to subtract a background due to the non-resonant contributions of the other elements of the sample. In the following we describe the procedure of this background subtraction for the example of the XMCD curves at the Cu-L$_{3,2}$-edges in the YBCO/LCMO superlattice. The same procedure has been used for the analysis of the XLD curves and also for the XMCD and XLD curves of the LSCO/LCMO multilayers.
\begin{figure*}[htb]
\includegraphics[width=\textwidth]{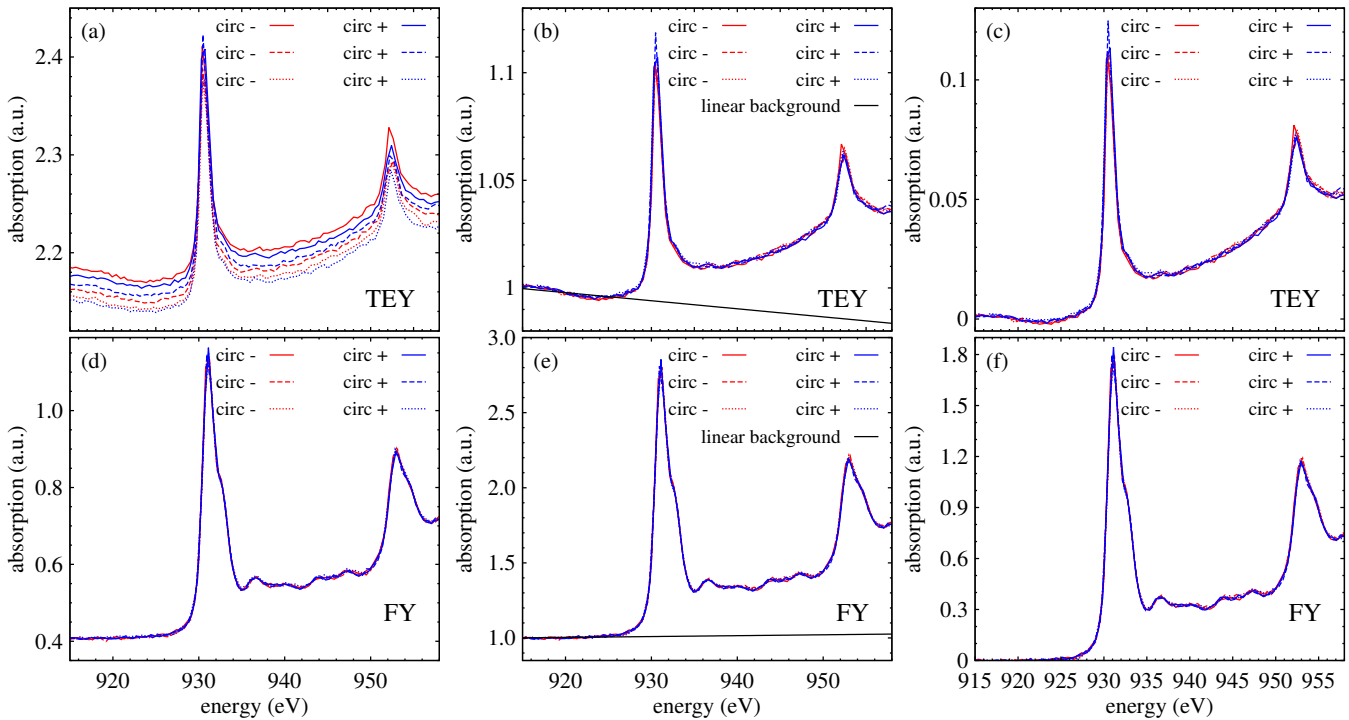}
\caption{(Color online) Illustration of the procedure for the background subtraction to obtain the resonant absorption at the Cu-L$_{3,2}$ edges in TEY (panels (a), (b) and (c)) and FY (panels (d), (e) and (f)) mode. Panels (a) and (b) show the raw data of the total electron and fluorescence yield for a series of three consecutive series of measurements at positive and negative helicity that were performed at 0.5\,T and 2\,K. Panels (b) and (e) display the same data after the normalization with respect to the intensity in the pre-edge region. The black line shows the extrapolation of the linear fit of the pre-edge region that is used to account for the background absorption. Panels (c) and (f) show the absorption at the Cu-L$_{3,2}$ edges obtained after the subtraction of this linear background.\label{Data_Cu}}
\end{figure*}

Panels (a) and (d) in Figure~\ref{Data_Cu} show the raw data for a series of XAS measurements which have been performed with an applied field of 0.5\,T at 2\,K for circular polarization with positive and negative helicity. For each polarization the measurement has been repeated three times. The TEY curves in panel (a) exhibit a noticeable intensity drift that is most likely caused by a weak charging effect of the sample. Panel (b) shows that this effect can be corrected for by normalizing the intensity with respect to the average value in the pre-edge region. Panels (d) and (e) show the corresponding correction procedure for the FY data where the intensity drift is significantly smaller. After this normalization the curves match very well in the regions before and after the edges and can now be directly compared.

The next step involves the subtraction of the background that arises from the non-resonant absorption due to the other elements in the sample. In a first approximation, the contribution of these remote absorption edges has been approximated with a linear function. Panels (b) and (d) show the result of the linear fit to the data in the pre-edge region. The TEY and FY curves after the subtraction of this linear background are displayed in panels (c) and (f). For the FY curve in panel (f) this yields a satisfactory result. For the TEY curve in panel (c) it is evident that this linear background subtraction is not as accurate. This is due to the very low probe depth of the TEY mode and the circumstance that the YBCO layer is buried below the topmost LCMO one. The Cu signal is therefore not much stronger than the background which is dominated by  post-edge features of the strong La-M$_{4,5}$ absorption edge near 850\,eV. For the sum-rule analysis of the spin and orbital moments we have therefore performed an additional background subtraction that is outlined in Ref.~\onlinecite{Uribe_2014a}.

Finally, for the Cu-L$_{3,2}$ edge FY and TEY curves shown in the main text, we have normalized the intensities with respect to the maximum of the peak at the L$_3$ edge. The $\mu_+$ and $\mu_-$ curves represent the average of the absorption curves for x-ray angular momentum antiparallel (positive helicity) and parallel (negative helicity) to the applied magnetic field, respectively, as measured for positive and negative field directions. This reduces the noise level and helps to remove instrumental artifacts. We note that the XTreme beamline allows one to change the polarization state of the x-rays (via a modification of the undulator shift) without affecting the beam path to the sample and thus without modifying the offset or the dispersion of the energy scale~\cite{Piamonteze_2012}.
\section{Weight of the peaks in the polarization-averaged absorption}\label{weight}
To relate the fitted intensity of the peaks with the fraction of the probed Cu atoms that contribute to each component in the absorption, we have integrated the polarization-averaged absorption, $2\mu_{ab}+\mu_c$, for each peak. In the averaged absorption all the anisotropic contributions are canceled. Its area is directly proportional to the hole density in the Cu-3d band, which has been set to $\underline{n}_{3d}=1$, and the number of the Cu atoms contributing to the particular transition~\cite{Stohr_1995}:
\begin{align}
  2A_{ab,i}+A_{c,i}&=CN_i\left[2(2\underline{n}_{3z^2-r^2}+6\underline{n}_{x^2-y^2})+8\underline{n}_{3z^2-r^2}\right]\nonumber\\
&=12CN_i.
\end{align}
Here $A_{\alpha,i}$ is the fitted area of peak $i$ for polarization in the direction $\alpha$, $N_i$ is the number of probed Cu atoms where the transition corresponding to peak $i$ occurs, $\underline{n}_j$ is the hole density in orbital $j$, and $C$ is an unknown proportionality constant. The different sensitivity to the $e_g$ orbitals for different polarizations arises from the dependence of the transition probabilities on the anisotropy of the charge distribution~\cite{Stohr_1995}. The normalized weight of each peak in the polarization-averaged absorption is therefore directly related to the fraction of the probed Cu atoms that contribute to the transition:
\begin{equation}
w_i=\frac{ 2A_{ab,i}-A_{c,i}}{\sum_{j}( 2A_{ab,j}-A_{c,j})}=\frac{12CN_i}{12C\sum_j N_j}=\frac{N_i}{\sum_j N_j},
\end{equation}
where the summation runs over the peaks contributing to the absorption and thus $\sum_j N_j$ corresponds to the total number of probed ions.


%

\end{document}